\documentclass[letterpaper, 10 pt, conference]{ieeeconf}  

\IEEEoverridecommandlockouts                              

\usepackage{epsfig} 
\usepackage{mathptmx} 
\usepackage{amsmath} 
\usepackage{cite}
\usepackage{url}
\usepackage{multirow}

\title{\LARGE \bf
Head--related Impulse Response Cues for Spatial Auditory Brain--computer Interface}

\author{Chisaki Nakaizumi$^{1}$, Shoji Makino$^{1}$, and Tomasz M. Rutkowski$^{1,2,3,*}$
\thanks{$^*$This research was supported in part by the Strategic Information and Communications R\&D Promotion Program (SCOPE) no. 121803027 of The Ministry of Internal Affairs and Communication in Japan.}
\thanks{$^{1}$Chisaki Nakaizumi, Shoji Makino and Tomasz M. Rutkowski are with Life Science Center of TARA and Department of Computer Science,
        University of Tsukuba, 1-1-1 Tennodai Tsukuba Ibaraki, Japan {\tt\small tomek@bci-lab.info} \qquad {\tt\small http://bci-lab.info/}}
\thanks{$^{2}$Tomasz M. Rutkowski is also with RIKEN Brain Science Institute, Wako-shi, Japan}
\thanks{$^{3}$Tomasz M. Rutkowski is the corresponding author.}     
\thanks{$^{*}$Tomasz M. Rutkowski was supported in part by YAMAHA Corporation.}     
        }

\begin{document}
\maketitle
\thispagestyle{empty}
\pagestyle{empty}

\begin{abstract}
This study provides a comprehensive test of a head--related impulse response (HRIR) cues for a spatial auditory brain--computer interface (saBCI) speller paradigm. We present a comparison with the conventional virtual sound headphone--based spatial auditory modality. We propose and optimize the three types of sound spatialization settings using a variable elevation in order to evaluate the HRIR efficacy for the saBCI. Three experienced and seven naive BCI users participated in the three experimental setups based on ten presented Japanese syllables. The obtained EEG auditory evoked potentials (AEP) resulted with encouragingly good and stable P300 responses in online BCI experiments. Our case study indicated that users could perceive elevation in the saBCI experiments generated using the HRIR measured from a general head model. The saBCI accuracy and information transfer rate (ITR) scores have been improved comparing to the classical horizontal plane--based virtual spatial sound reproduction modality, as far as the healthy users in the current pilot study are concerned.
\end{abstract}

\section{Introduction}

BCI is a technology that uses brain neuronal signals to operate a computer without any muscle movements. Therefore, it is expected to provide a speller for disabled people such as patients suffering from the amyotrophic lateral sclerosis (ALS)~\cite{kubler2009brain}. Although currently a successful visual BCI modality could provide a fast speller, the advanced patients who are in a locked--in state cannot use it because they lose any intentional muscle control including even eye blinks~\cite{bciBOOKwolpaw}. Auditory BCI can be an alternative method because it does not require a good sight or eye movements~\cite{kubler2009brain,iwpash2009tomek,bciSPATIALaudio2010,tomekEMBC2011,MoonJeongBCImeeting2013}. 
We propose an alternative method to extend the previously published by our group spatial auditory BCI (saBCI) paradigms~\cite{iwpash2009tomek,MoonJeongBCImeeting2013} by making use of a head related impulse response (HRIR) for the virtual spatial sound images reproduction with headphones. 
Our research target is the virtual sound saBCI using the HRIR--based spatial cues to create the non--invasive and auditory stimulus--driven paradigm, which does not require a long training. 
HRIR appends interaural--intensity--differences (IID), interaural--time--differences (ITD), and spectral modifications to create the spatial stimuli, while a vector--based amplitude panning (VBAP) appends only the IID. The HRIR allows for more precise and fully spatial virtual sound images positioning utilizing even the user not own HRIR measurements~\cite{book:auditoryNEUROSCIENCE}.

In our previous study~\cite{nakaizumi2013HRIR} we evaluated saBCI feasibility with the HRIR--based spatial sound generator, and compared it with a formerly reported vector--based--amplitude--panning (VBAP)--based spatial auditory experiments~\cite{MoonJeongBCImeeting2013}. The above study used only five Japanese vowels distributed horizontally every $40^{\circ}$ in front of a subject head. The pilot study resulted with clear P300 responses and it has shown that HRIR modality could improve spelling accuracy comparing to the conventional spatial sound generation methods.
In the study presented in this paper we introduce ten Japanese kana syllables--based speller using the sound elevation features as the second step. We also compare various sound elevation settings.
However, it is usually difficult to precisely perceive various sound elevations using not the user own HRIR, because the elevation perception is highly individual and it is influenced by the shape of an auricle~\cite{book:auditoryNEUROSCIENCE}.
Unfortunately, it would be very difficult to record a bedridden ALS--patient own HRIR.
Therefore, we propose to test the HRIR--based saBCI efficacy using a general head model (KEMAR).
We conduct psychophysical and EEG experiments in online saBCI ten Japanese kana syllables spelling experiments in order to test our research hypothesis of the KEMAR HRIR--based paradigm practical feasibility. 

From now on the paper is organized as follows: next section describes methods and experimental set up; 
third section reports results of the $P300$ response--based saBCI speller classification accuracies and information transfer rate (ITR) in comparison to a conventional method; finally, conclusions and future research directions summarize the paper.


\section{Materials and Methods}

\begin{figure}[!t]
	\centering
	\vspace{0.3cm}
	\includegraphics[width=\linewidth]{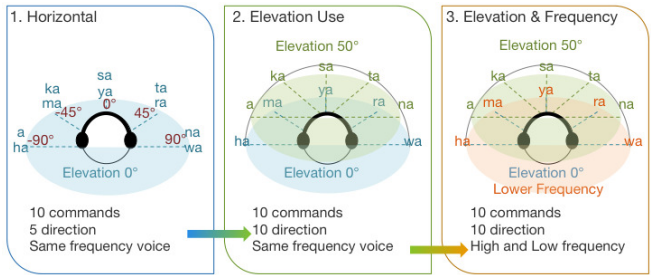}
	\caption{A diagram of the three proposed virtual spatial stimulus sound settings. The left box depicts the horizontal only placement--based setting; the middle the elevation use--based case; and the right box the elevation use with an additional frequency shifting set up.}\label{fig:3modalities}
	\vspace{-0.3cm}
\end{figure}

In order to prepare the saBCI stimuli to create an oddball paradigm generating the P300 responses~\cite{bciBOOKwolpaw}, all spatial sound images were created using HRIR of the general head model.
Ten Japanese syllables with a vowel "a" were selected for as sound stimuli in this project.
The spoken syllables were taken from a public Japanese sound dataset of a female speech~\cite{amano2009development}. 
The monaural sound stimuli were spatially distributed using a public domain \textsc{KEMAR's HRTF database} provided by the MIT Media Lab~\cite{MITkemarHRTF}. 
In order to generate a stereo sound placed at a spatial location at an azimuth of $\theta$ and an elevation of $\phi$ the following procedure was applied. Let $h_{l,\theta,\phi}$ and $h_{r,\theta,\phi}$ be the minimum--phase impulse responses from the \textsc{KEMAR HRTF database} measured at the chosen azimuth $\theta$ and the elevation $\phi$ at the left ($l$) and right ($r$) ears. 
The stereo spatial sound delivered via headphones to the left and right ears respectively could be constructed, in time domain using HRIR, as a two--dimensional signal composed of the left $x_l(t)$ and right $x_r(t)$ headphone channels as follows,
\begin{align}
	x_l(t) = \sum^{n-1}_{\tau=0}h_{l,\theta,\phi}(\tau)s(t-\tau),\label{eq:HRIR} \\ 
	x_r(t) = \sum^{n-1}_{\tau=0}h_{r,\theta,\phi}(\tau)s(t-\tau),\nonumber
\end{align}
where $\tau$ denotes sample time delay and $n$ is the HRIR length as obtained from the \textsc{HRTF Database}~\cite{cipicHRTF}.
The so created spatial acoustic stimuli were delivered to the left and right ears of the user through the ear--fitting portable headphones \textsc{SENNHEISER CX 400II}. 
Three proposed spatial settings to evaluate the feasibility of the sound elevation using the HRIR for the saBCI are presented in Figure~\ref{fig:3modalities}. 
The first spatial stimulus setting included only the horizontal sound images' placement of the ten saBCI commands. The sound images were distributed at five directions every $45^\circ$ on the horizontal plane at the user ears' level. Two types of commands were delivered from the same spatial directions.
The second stimulus setting included the sound elevation variability.
Ten stitmuli were distributed among the all different directions.
Five commands were localized at elevation of $50^{\circ}$ from the user ears' level and every $45^{\circ}$ horizontally.
The third setting included also the variable elevation and an additional option of a frequency modulation  appended to discriminate the sound images originating from various elevations.
In the psychoacoustics there is a well--known property called a tonal bell causing a higher--frequency sound to be perceived as originating from a higher elevation~\cite{book:auditoryNEUROSCIENCE}. To simulate the above effect we shifted in frequency domain the stimuli, which shall be perceived at higher elevations, using the \textsc{TANDEM-STRAIGHT} method~\cite{tandemSTRAIGHT}. The sound stimuli to be perceived at elevation of $50^{\circ}$ had shifted up their fundamental frequencies by $24$~Hz.


All of the experiments reported in this paper were performed in the Life Science Center of TARA, University of Tsukuba, Japan. Ten healthy users participated in our study. They were comprised of seven naive and three experienced BCI users. The average age was of $24.7$ years old (standard deviation $6.48$ years old, three males and seven females).
The psychophysical and online EEG saBCI experiments were conducted in accordance with \emph{The World Medical Association Declaration of Helsinki - Ethical Principles for Medical Research Involving Human Subjects}. The experimental procedures were approved and designed in agreement with the ethical committee guidelines of the Faculty of Engineering, Information and Systems at University of Tsukuba, Japan.

The psychophysical experiments were conducted to examine a perception of elevation and preferences for each spatial sound setting. The users were instructed to respond by pressing the button as soon as possible after they perceived the \emph{target} stimulus as in the classical oddball paradigm~\cite{bciBOOKwolpaw}. In a single experimental session $10$~\emph{targets} and $90$~\emph{non-targets} were presented. Each experiment was comprised of three sessions for every spatial sound setting. The stimulus duration was set to $300$~ms and the inter--stimulus--interval (ISI) to $700$~ms.
The online EEG experiments were conducted to investigate whether P300 responses could be evoked in the various spatial sound settings and to compare the saBCI classification accuracies, as well as an efficaty of each set up.
The brain signals were collected by a bio--signal amplifier system \textsf{g.USBamp} by g.tec Medical Engineering GmbH, Austria. The EEG signals were captured by sixteen active gel--based electrodes \textsf{g.LADYbird} attached to the following head locations \emph{Cz, Pz, P3, P4, Cp5, Cp6, P1, P2, Poz, C1, C2, FC1, FC2,} and \emph{FCz} as in the extended $10/10$ international system~\cite{bciBOOKwolpaw}. The ground electrode was attached on the forehead at the \emph{FPz} location, and the reference  on the  user's left earlobe respectively. An in--house extended \textsc{BCI2000}~\cite{bci2000} software was used for the saBCI experiments to present stimuli and display online classification results. 

A single EEG experimental session was comprised of one training and two test runs for each spatial stimulus setting. 
In the training run, EEG brainwaves were recorded and next classifier parameters were calculated. In order to spell a single character, $15$~\emph{targets} and $135$~\emph{non-targets} were presented for ERP response averaging. 
In the test runs, user spelled two sets of the three words that the covered ten \emph{targets} (\emph{a--ta--ma; ha--ra;  sa--wa--ya--ka--na; wa--ta; ha--na--ya; a--ka--ra--sa--ma}). 
For spelling single character, $5$~\emph{target} and $45$~\emph{non-target} stimuli were presented continuously as for single syllable (five ERPs averaging scenario). Experiment order of every spatial sound setting and spelled words were randomized for each subject.
The sound stimulus duration was set to $300$~ms and the inter--stimulus--interval (ISI) to $150$~ms.
The EEG sampling rate was of $512$~Hz and a notch filter to remove electric power lines interface of $50$~Hz was applied in a rejection band of $48\sim52$~Hz. 
The band--pass filter was set at $0.1$~Hz and $60$~Hz cutoff frequencies. The acquired EEG brain signals were classified online by the in--house extended \textsc{BCI2000} application using a stepwise linear discriminant analysis (SWLDA)~\cite{krusienski2006} classifier with features drawn from the $0\sim800$~ms ERP interval, with removal of the least significant input features, having $p > 0.15$, and with the final discriminant function restricted to contain a maximum of $60$ features.
The participating users answered also questionnaires, asking about which modality they preferred, after finishing psychophysical and EEG experiments.

\section{Results}
\begin{figure}[t]
	\centering
	\includegraphics[width=0.9\linewidth]{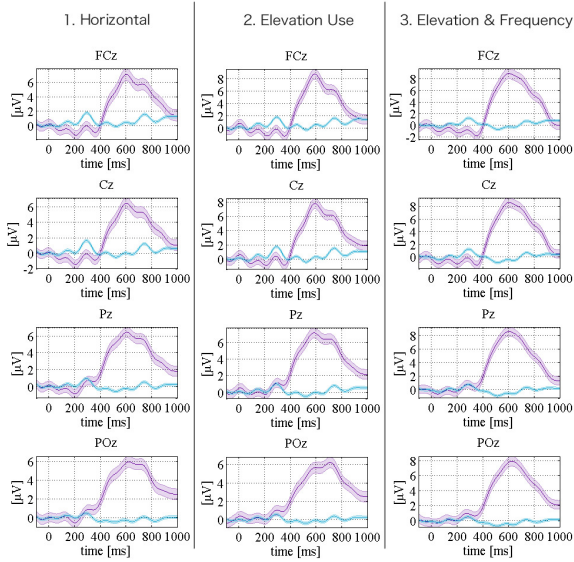}
	\caption{Grand mean averaged EEG ERP responses together with standard error bars for all the participating users in the study plotted for representative electrodes separately. Purple lines depict the grand mean averaged \emph{targets} with clear P300 responses, while blue traces are the \emph{non--targets.} Eye blink artifacts were removed with $80~\mu V$ thresholding.}\label{fig:EEGERP}
	\vspace{-0.3cm}
\end{figure}
\begin{figure}[t]
	\centering
	\includegraphics[width=\linewidth]{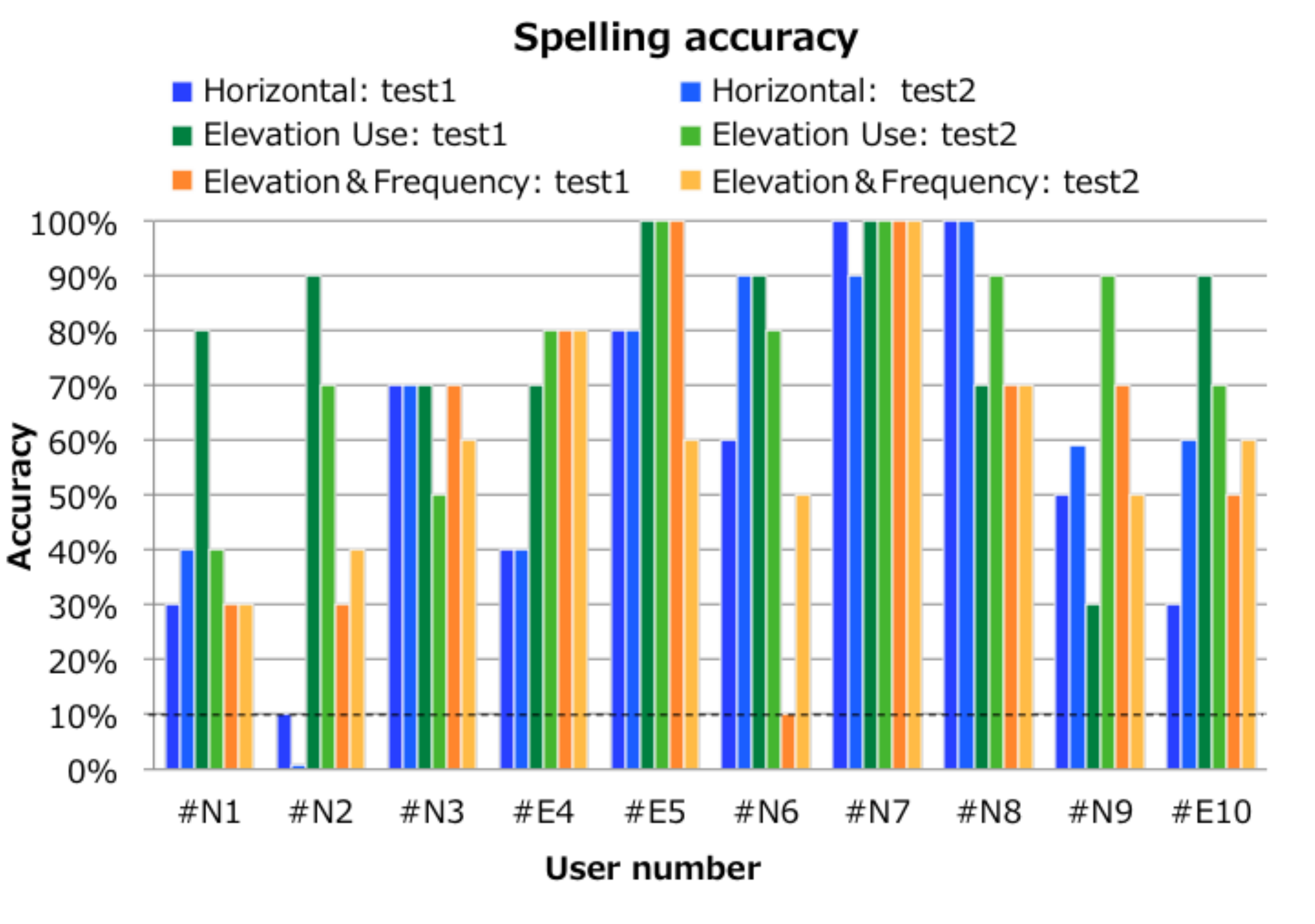}
	\caption{The average saBCI spelling accuracies of all the users.
The chance level was of $10\%$. The blue bars depict the saBCI accuracy of horizontal sound images placement; green bars represent elevation--based  results; and the orange color the elevation and frequency settings, respectively. Th users $\#4, \#5$ and $\#10$ were the experienced BCI users, while the remaining were the naive participants.}\label{fig:SpellingAccuracy}
\vspace{-0.3cm}
\end{figure}
\begin{figure}[tb]
	\vspace{0.2cm}
	\centering
	\includegraphics[height=6.2cm]{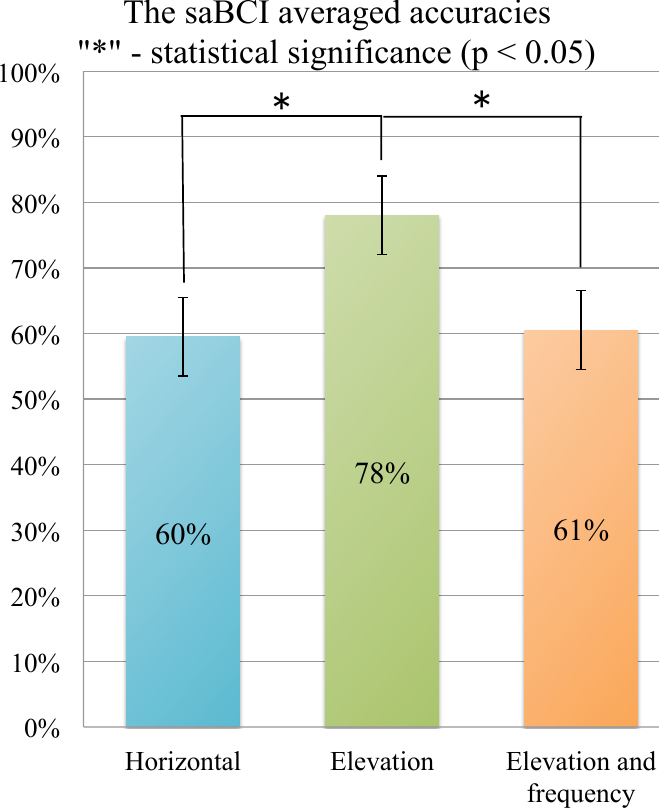}
	\includegraphics[height=5.9cm]{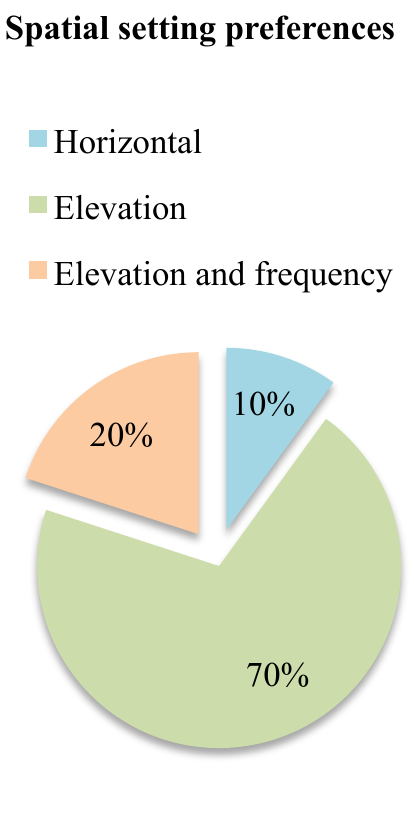}
	\caption{This averaged saBCI spelling accuracies for each of the tested spatial sound settings.
The theoretical chance level was of $10\%$. The blue color depicts the accuracy of horizontal sound setting; green the elevation variability; and orange the elevation and frequency case, respectively. The right side pie chart depicts the user preferences of the three spatial sound settings with the same color coding, respectively.}\label{fig:AveragedAccuracy}
	\vspace{-0.5cm}
\end{figure}

%
%
In the psychophysical experiments the users could perceive the various elevation settings. The questionnaire answers indicated that the frequency shifting supported the stimulus discrimination.
The elevation and frequency variable setting resulted with the best perception of the spatial sound image stimuli.
Four out of six participants preferred elevation and frequency--based spatial sound setting.
The results of the saBCI EEG experiment are depicted in Figure~\ref{fig:EEGERP}.
Each column presents the grand mean averaged ERP results at representative four electrodes for the three proposed spatial sound settings. We confirmed the clear P300 responses in latency ranges of $400\sim1000$~ms  and their usability for the subsequent classification.
Figure~\ref{fig:SpellingAccuracy} presents the classification accuracies of the P300 responses as obtained with the SWLDA classifier. The theoretical chance level was of $10$\% in this study. 
All users scored with accuracies above $70$\% at the best. There were three users who resulted with $100\%$ accuracies as the best in the reported experiments. 
The average scores were obtained as mean values calculated from the two saBCI test spelling sessions of all the users (training sessions were not included in the saBCI accuracy calculations). The results are shown in Figure~\ref{fig:AveragedAccuracy}. There were significant differences between horizontal and elevation use; elevation only versus elevation and frequency spatial sound settings as calculated with the t--test (the significant level was $p < 0.05$). The $70$\% of the users preferred elevation--based spatial sound setting. 
%
The important outcome of the presented study was that the users could clearly perceive elevation variability using not their own HRIR filters.
Although the results of psychophysical experiments have shown that the users preferred elevation and frequency spatial sound setting, the online saBCI accuracies suggested that P300 responses generated in elevation variability setting where the best.
In order to compare the developed HRIR--based saBCI we calculated the information transfer rate (ITR) scores considered as a major comparison measure among the BCI paradigms~\cite{bciSPATIALaudio2010}. ITR scores were calculated as follows:
\begin{align}\label{eq:ITR}
	ITR &= V\cdot R,&  \\
 	R &= \log_2 N + P\cdot \log_2 P + (1-P)\cdot \log_2\left(\frac{1-P}{N-1}\right),&\nonumber
\end{align}
where $V$ was the classification speed in selections/minute; $R$ represented the number of bits/selection; $N$ was the number of classes ($10$ in this study), and $P$ the classification accuracy.
The conditions contributing to ITR were in a trade--off relationship with the task easiness. 
For example, the short ISI could improve the ITR, but it could cause the task to be difficult.
ITR could increase with larger number of commands, higher accuracies, shorter ISI, and a smaller number of the averaged trials.
We also compared the ITR scores of the three proposed modalities with our previous project results and with the  vector-based-amplitude-panning (VBAP)--based spatial auditory approach, which was regarded as a conventional method~\cite{MoonJeongBCImeeting2013}. The VBAP experiment was conducted for two sessions and with $16$~BCI--naive users~\cite{MoonJeongBCImeeting2013}. The electrode positions were the same as in the  experiments reported in this paper. The sound stimuli were presented with small ear--fitting headphones (\textsc{SENNHEISER CX 400II}) in the all the studies.  
The spatial locations of the stimulus sound images, the number of commands, and ISI settings are summarized in Table~\ref{tab:ITRtable}.
The ITR scores of the elevation use spatial setting reached $14.92$~bit/min. The other proposed modalities scored above $8.5$~bit/min and also exceeded our previous research results~\cite{nakaizumi2013HRIR}. The headphone--based virtual saBCI took one step forward.
Although the HRIR based saBCI resulted with better accuracy scores, this results cannot be compared for a statistical significance with the former study due to the different subject groups.
\begin{table}[b]
	\centering
	\caption{The averaged accuracies and ITRs of the proposed, our previous HRIR project and the conventional VBAP methods}
	\label{tab:ITRtable}
	\begin{tabular}{|l|c|c|c|}
	\hline 
	\multicolumn{2}{|c|}{\multirow{2}{*}{Spatial sound mode}} & \multicolumn{1}{|c|}{Averaged}& \multicolumn{1}{|c|}{ITR}\\
	\multicolumn{2}{|c|}{}& \multicolumn{1}{|c|}{accuracy}& \multicolumn{1}{|c|}{(bit/min)}\\ \hline\hline
The proposed method with 	& Horizontal			&$59.5$\%						&$~~8.51$	\\\cline{2-4}
the use of various HRIRs; 	& Elevation  					&$78.0$\%	& $14.92$	\\\cline{2-4}
$10$~commands; ISI=$150$~ms						& Elev. and freq.  	&$60.5$\%			&$~~8.81$	\\ \hline
Previous HRIR		& \multirow{2}{*}{Horizontal}	& \multirow{2}{*}{$55.8$\%} 	& \multirow{2}{*}{$~~1.79$}	\\
$5$~commands; ISI=$300$~ms&					&		&		\\ \hline	
Conventional VBAP;	& \multirow{2}{*}{Horizontal}			& \multirow{2}{*}{$45.6$\%}	&\multirow{2}{*}{$~~0.57$}	\\
$5$~commands; ISI=$500$~ms	&					&		&		\\ \hline
	\end{tabular}
\end{table}
We also analyzed spelling accuracy for the various horizontal azimuths as shown in Table~\ref{tab:CMX}.
Compared with the horizontal placement modality, the remaining two spatial sound settings resulted with reduced accuracy errors.
The results have shown that the variable stimuli elevation helped the users to distinguish better the horizontal sound locations either.
\begin{table}[t]
	\vspace{0.2cm}
	\centering
	\caption{The mean spelling accuracy for various horizontal azimuths}\label{tab:CMX}
	\begin{tabular}{|l||c|c|c|c|c|}
	\hline 
	\multicolumn{1}{|c||}{\multirow{2}{*}{Sound mode}}
      & \multicolumn{5}{|c|}{Sound azimuth--based saBCI mean accuracies}
      \\ \cline{2-6}
	& $-90^{\circ}$	& $-45^{\circ}$	& $0^{\circ}$	& $45^{\circ}$	& $90^{\circ}$ \\ \hline \hline
Horizontal 	&$85\%$	&$80\%$		&$95\%$		&$90\%$		&$90\%$	
\\  \hline
Elevation	&$90\%$	&$95\%$		&$95\%$		&$90\%$		&$95\%$	
\\ \hline
Elevation and & \multirow{2}{*}{$90\%$} & \multirow{2}{*}{$95\%$}		& \multirow{2}{*}{$90\%$} & \multirow{2}{*}{$85\%$}	& \multirow{2}{*}{$100\%$} 
\\
frequency & & & & &\\	 \hline
	\end{tabular}
	\vspace{-0.2cm}
\end{table}

\section{Conclusions}

The presented EEG results confirmed the P300 responses feasibility among the experienced and naive saBCI users.
The proposed spatial sound stimulus settings for various elevations obtained using HRIR were effective for saBCI paradigm. Additionally, the short ISI did not distract the users' perception, rather it apparently sharpened it resulting with the better saBCI accuracies.

The obtained ITRs resulted with better scores comparing to our previous study using simple HRIR and VBAP--based spatial sound virtualization.

Nevertheless, current study is not yet ready to compete with the faster visual BCI spellers. Furthermore, it is necessary to improve the ITR for a comfortable online saBCI--based spelling. 
We plan to extend the proposed saBCI to realize the full set Japanese kana characters--based speller, and to design a more effective spatial sound placement using more precise elevation settings.

\addtolength{\textheight}{-12cm}   


\bibliographystyle{IEEEtran}
\bibliography{chisaki}

\end{document}